# Compression behavior of simply-supported and fully embedded monolayer graphene: theory and experiment

Emmanuel N. Koukaras[1], Charalampos Androulidakis[1,2], George Anagnostopoulos[1], Konstantinos Papagelis[1,2] and Costas Galiotis[*,1,3]

[1]Institute of Chemical Engineering Sciences, Foundation of Research and Technology-Hellas (FORTH/ICE-HT), Stadiou Street, Platani, Patras, 26504 Greece

[2]Department of Materials Science, University of Patras, Patras, 26504 Greece

[3]Department of Chemical Engineering, University of Patras, Patras 26504 Greece

[*]Corresponding author: c.galiotis@iceht.forth.gr or galiotis@chemeng.upatras.gr

## ABSTRACT

Single layer graphene simply-supported on a polymer substrate was subjected to axial compression and its behavior upon loading was monitored with laser Raman spectroscopy (LRS). The graphene was found to fail by wrinkling (buckling) at a critical strain of −0.30% and at a compressive stress of ~1.6 GPa, as revealed by the conversion of the spectroscopic data to actual stress-strain curves. This contrasts with the value of –0.60% and stress of ~3.8 GPa required for failure initiation in the fully embedded case. To elucidate the failure mechanisms in the two cases examined, molecular dynamics simulations employing the AIREBO potential were performed. We assess the impact of surface roughness, graphene–polymer interaction, and of thermal (phonon) ripples on the onset of wrinkle formation. Overall good agreement was found between theory and experiment. As argued herein, the understanding and control of out-of-plane phenomena upon mechanical loading of graphene are important prerequisites for the design and function of new graphene-based devices.

## 1. Introduction

Graphene is a two dimensional crystal with thickness of one carbon atom forming a hexagonal honeycomb ordered structure. It is the thinnest known material exhibiting unparalleled stiffness of 1 TPa when it is completely flat, high extensibility and fracture strength which can be as high as ~130 GPa[1]. These mechanical properties[2] in tandem with its remarkable electronic properties — such as having an electron–acoustic phonon scattering mean free path in the μm order, and recorded room temperature carrier mobility up to ~$4\times10^4$ cm$^2$V$^{-1}$s$^{-1}$ that can possibly reach up to $2\times10^5$ cm$^2$V$^{-1}$s$^{-1}$ once effects of substrate (extrinsic scattering) are removed or the influence of flexural phonons is suppressed (for example through application of tension)[3-7] — make graphene a potential candidate for various applications such as sensors, flexible electronics and as a reinforcing filler in nano-composites. Crucial to these applications is the knowledge of the deformation mechanism of the graphene under mechanical loadings either resting on a substrate or fully embedded in polymer matrices.

Upon compressive loading graphene undergoes wrinkling/buckling type of instabilities because of its initial very low bending rigidity[3]. Although these instabilities constitute an elastic Euler-type (geometric) failure, harnessing them provides such control that allows for intriguing phenomena to be induced by design such as the transformation of graphene to an auxetic material[8,9]. Other forms of instabilities have also been observed such as crumpling under biaxial compression[10], rippling under tension[11], and origami patterns of graphene paper adhered on a pre-stretched elastomeric substrate[12].

Previous theoretical works that include folding[13] and crumpling[14] of graphene on a substrate have been examined using an atomistic-based continuum approach and molecular mechanics respectively. Graphene on a substrate has been modelled analytically as a hexagonal 2-lattice and studied using continuum mechanics[15,16]. Other effects such as the sequential period-doubling bifurcations for graphene bonded to a PDMS substrate were also studied numerically[17]. Moreover, the buckling of a monolayer graphene on an oxide substrate has been studied theoretically for the cases where the morphology of the substrate is flat or corrugated[18]. Other aspects such as the wet adhesion of the graphene or mechanics related have also been examined using molecular dynamics simulations[19-21].

Most experimental procedures for subjecting a monolayer graphene (or any 2-D crystal) to strain gradients involve the use of polymer substrates combined with Raman spectroscopy[2,22,23,24]. The graphene is placed on the surface of a polymer beam and by bending[22, 23] (or stretching[24]) axial strain is transmitted to graphene through shear at the graphene/polymer interface. Upon loading, Raman measurements are collected by monitoring the position of the graphene 2D and G peaks at various strain levels. The shifts of the Raman peaks upon mechanical loading provide information for the state of stress/strain in the graphene itself and a measure of stress-transfer efficiency[23]. Recently, the technique of flexed beams has been extended to two dimensions that allow the study of graphene and other 2-dimensional materials subjected to controlled biaxial tensile deformations[25].

In this work single layer graphene simply supported on a plastic bar, is examined experimentally using the approach described above. Wrinkle formation is observed at a compressive strain of −0.30% and theoretical modelling is used to elucidate and interpret these findings. Moreover, based on the methodology reported previously[26,27] the spectroscopic data are converted to axial stress/strain data. The experimental results for graphene resting over polymers under compression are complemented by molecular dynamic simulations. The results also revealed the influence of graphene's intrinsic thermal (phonon) fluctuations to the wrinkling formation. Good agreement is found between experiments, theory and simulations.

## 2. Materials and methods

### 2.1 Experimental

A plastic bar of PMMA (poly-methylmethacrylate) was used as the substrate. A thin layer of SU-8 photoresist with thickness of ~200 nm was spin coated on the PMMA to improve the optical contrast of the graphenes. Exfoliated graphitic materials (from HOPG) were deposited on the PMMA/SU-8 substrate using the scotch tape method and appropriate flakes were located with an optical microscope. The exact thickness of the graphene was identified by the corresponding spectra of the 2D Raman line. A four-point-bending jig[28] which is placed under a Raman microscope was used for subjecting the samples to compression. The strain was applied incrementally with a step of −0.05% and at every strain level the Raman spectra for the 2D and G peaks were recorded *in situ*. The Raman spectra measured at 785 nm and the laser power was kept below 1 mW in order to avoid local heating of the samples.

### 2.2 Details on the computations

The molecular dynamics simulations have been performed employing the AIREBO potential[29] to model carbon–carbon interactions in graphene. The environment of graphene is modelled through combined mathematical surfaces with adjustable interaction with the carbon atoms of graphene. Initial surfaces are mathematically ideal and imperfections are modelled as required. This approach allows for the introduction and individual study of the effects of imperfections. Further details are provided in the corresponding sections. All of the simulations have been performed using a large computational cell of 852.0×196.8 Å$^2$ and periodic boundary conditions, with a graphene consisting in total of 64000 carbon atoms, that allows to capture all of the effects of interest. A time step of 1.0 fs was used throughout. The compression of graphene was strain based and performed with a low constant engineering strain rate[30] of 5×10$^{-3}$ %/ps (in the order of the lowest used in the literature [31,32]). All of the molecular dynamics simulations were performed using the LAMMPS package[30]. Optical inspection of the trajectory frames was performed using the Ovito package[33].

## 3. Results and Discussion

### 3.1 Embedded graphene under compression

In previous work[28] the response of monolayer graphene embedded in a polymer matrix under compression was studied experimentally and theoretically. It was found that buckling failure initiates at compressive strains with a mean value of –0.60%. From results obtained by analytical modeling through Euler mechanics combined with a Winkler approach a wrinkle wavelength of the order of 1–2 nm was estimated. Here we try to understand in-depth the instability mechanism of the embedded graphene, as well as the origin of the non-linear mechanical behavior by theoretical modeling and further assisted by molecular dynamics simulations. The compression instability of mono-layer graphene embedded in polymer matrices is modeled using the Winkler's approach [28]. In this approach the interaction between the graphene and the polymer was simulated with linear elastic springs. The following set of equations is required to address the problem:

$$\left. \begin{array}{l} \varepsilon_{cr} = \pi^2 \dfrac{D}{C}\dfrac{k}{w^2} + \dfrac{l^2}{\pi^2 C}\left(\dfrac{K_W}{m^2}\right) \\ k = \left(\dfrac{mw}{l} + \dfrac{l}{mw}\right)^2 \\ m^2(m+1)^2 = \dfrac{l^4}{w^4} + \dfrac{l^4 K_W}{\pi^4 D} \end{array} \right\} \text{Winkler's model ,}$$

where $\varepsilon_{cr}$ is the critical strain for buckling instability, $D$ and $C$ are the flexural and tension rigidity of the graphene respectively, $l$ and $w$ are the flake's dimensions, $k$ is a geometric term given by the second equation, $m$ is the number of the half-waves that the plate buckles and is estimated by the third equation, and $K_W$ is the Winkler's modulus.

The wrinkle wavelength is given by

$$\lambda = \frac{l(1-\varepsilon_{cr})}{m},$$

where $l$ is the length of the flake.

The Winkler modulus that was estimated by DFT analysis is $K_W$ = 21.77 GPa/nm for a PMMA/graphene system and corresponds to a compressive critical strain of −1.1%.[28] The experimentally obtained value for the mean critical strain is $K_W$ ~6 GPa/ nm which is lower from the value of DFT.[28] This is not surprising since the DFT results represent the ideal interaction between polymer and graphene.

We performed molecular dynamics simulations of monolayer graphene embedded between two surfaces with which graphene's carbon atoms interact. The form of the interaction used is a 12–6 Lennard–Jones potential, $V(r) = 4\varepsilon\left[(\sigma/r)^{12} - (\sigma/r)^{6}\right], r \leq r_c$, with suitable values for the parameters. To establish these parameters we consult the few existing values from the literature. Lv et al. studied[34] interfacial bonding characteristics between graphene and PMMA and polyethylene (PE) polymer matrices. They embedded a graphene sheet in the polymer matrices and amongst other quantities they calculated the interfacial bonding energy (surface energy) $\gamma$, defined as half the total interaction energy (per area). For graphene fully embedded in PMMA they find a $\gamma$ value of –6.82 meV/atom (–0.06 kcal mol$^{-1}$Å$^{-2}$). Energy conversions are straightforward using the area per carbon atom in graphene, $A_{atomic} = 1/2|\mathbf{a}_1 \times \mathbf{a}_2| = 1/2 a^2 \sin 60° = 2.62 Å^{-2}$, where $a$ is the lattice constant of graphene. Since these values for $\gamma$ were calculated from total interaction energies for embedded graphenes, i.e. two interfaces or four contacting surfaces, the corresponding interaction energies per area per interface is $2\gamma$. These values are somewhat higher than those reported previously by Androulidakis et al. [28] which were obtained from calculations based on high accuracy dispersion corrected density functional theory. In that work, depending on the configuration of PMMA, the reported interaction energies range from 6.79–9.23 meV/atom (0.25–0.35 kcal mol$^{-1}$ Å$^{-2}$). A contributing factor to the differences in values is an overestimation that may result from the interaction (captured by the COMPASS potential[35]) of the hydrogen atoms used in the work of Lv et al. for the passivation of terminal carbon atoms in graphene with the carbonyl oxygens of the PMMA monomers. This is further supported by the significant increase reported therein of the (magnitude of the) interfacial bonding energy prior to the full withdrawal of the graphene from the polymer matrix in the pullout simulation with unfunctionalized graphene [34]. Considering the differences between the two levels of theories and the type of modeling, the agreement between the two is noteworthy. The parameters used for the Lennard–Jones potential are $\sigma$ = 3.133 Å with a cutoff of $r_c$ = 6.5 Å, throughout, and $\varepsilon$ = $10^{-3}$, $6.79 \times 10^{-3}$, $10^{-2}$, and $2 \times 10^{-2}$ eV

In **figure 1a** we show images in perspective of the graphene sample embedded between two interacting surfaces (the embedding surfaces are not shown for clarity). The simulations were performed at a temperature of $T$ = 300K. The distance between the two surfaces is $2Z_0$, with $Z_0$ = 3.7 Å being half of the intersurface distance where the graphene lies. The images correspond to levels of strain increasing from top to bottom. Initially no buckling is noted while at high strain levels (see **figures 1ab**) a systematic fluctuating pattern arises. At significantly higher strain levels bifurcation patterns also arise (shown in the third image in each of **figures 1ab**). We analyze several strain levels by producing average squared height plots that we show in **figure 2**. Based on these plots (region in which the slope of the average squared heights vs strain notably increases) along with optical inspection of the images, the critical strain for buckling is –0.75% to –1.26%. This range of values is somewhat higher than the corresponding experimental one. The main source for this discrepancy is that we have not yet accounted for imperfections of the surrounding surface, i.e. they are ideally smooth, and the surfaces are perfectly firm (they do not recede).

The results of **figure 1** confirm the theoretical assumption that the embedded graphene exhibits a sinusoidal form of buckling with multiple waves under compression. Also, good agreement is found between theory and simulations for the values of the buckling wavelength. This is a significant point because the form of failure cannot be observed optically due to the surrounding polymers and this seems to be the only viable way to assess the form of instability. Similar results from MD simulations have been obtained elsewhere[36] that also confirmed the theoretical results.

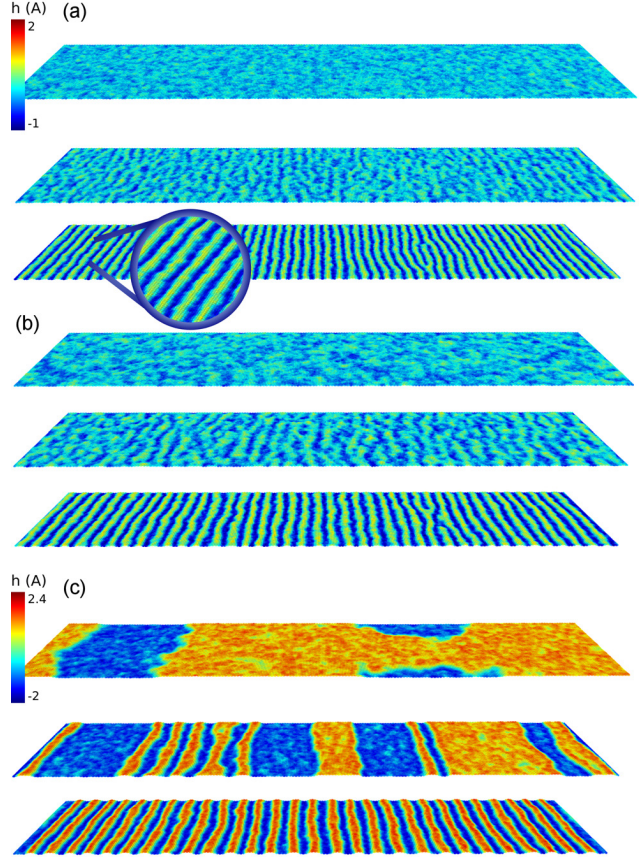

**Figure 1.** Compression stages for embedded graphene with different $Z_0$ values. The embedding surfaces are not shown. The values of $Z_0$ are (a) 3.7 Å, (b) 3.9 Å, and (c) 4.7 Å. Strain levels increase from top to bottom as (a) 0.0%, 1.56%, 3.66%, (b) 0.0%, 0.92%, 2.65%, and (c) 0.0%, 1.32%, 3.36%. The color scales is chosen for clarity, not for accuracy.

In **figure 2** we also provide the height plots with slightly increased distance $Z_0$. We find that small increases in $Z_0$ lead to significant changes in the response of graphene to compression. As shown in **figure 1b** and **figure 2,** by increasing $Z_0$ to 3.9 Å the strain level at which wave patterns form is significantly reduced from 1.56% to 0.92%. It is conceivable that such small differences are readily encountered in real embedded graphene samples and this sensitivity should be considered when interpreting experimental results. Further increase in $Z_0$ continues to change the behavior of graphene, to the point that when $Z_0 = 4.7$ Å the embedded graphene at its relaxed state (0.0% strain) exhibits extensive regions that adhere to either one of the surfaces, as shown in **figure 1c**. However, this case corresponds to a poorly embedded sample, which starts to occur at the smaller value of $Z_0 =$ 4.2 Å that corresponds to an overall ~1 Å departure from a perfectly embedded graphene (see **figure** 2 at zero strain level). Although we find from **figure 2** that the critical strain for buckling is significantly reduced compared to the ideal case, which for $Z_0 = 3.9$ Å is already in the range of –0.7% to –0.9%, i.e. in broad agreement the value of –0.6% found by Androulidakis *et al.*, this range is further reduced (without resorting to the poorly adhered cases) when alternate or supplementary mechanisms are considered, some of which are proposed in a following section.

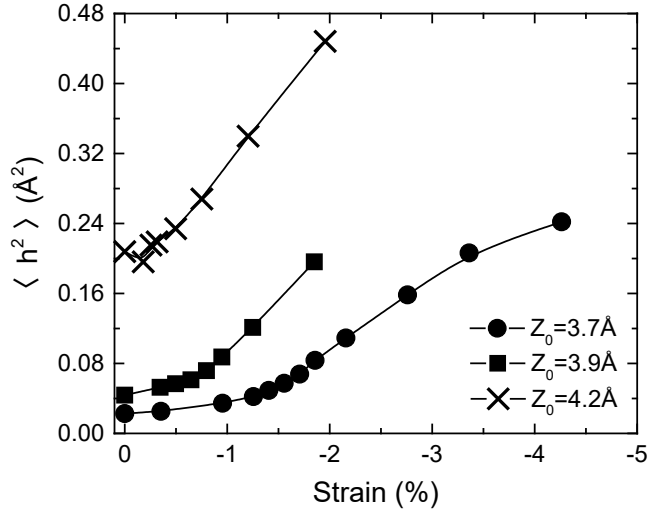

**Figure 2.** Squared height plots of the atomic distances from the central plane between the two embedding surfaces, for various intersurface distances $2Z_0$ (at $T = 300$K). The lines are guides to the eye.

The simulations provide valuable information for the origin of the non-linearity that it is observed in the compression stress-strain behavior of embedded monolayer graphene [27,28]. From **figures 1** and **2** it is deduced that there is a gradual increase on average of the atomic distances from the central plane (between the two embedding surfaces), as the assembly is compressed, that reaches a critical threshold value per applied strain when a wrinkling pattern is formed. This interface weakening with strain gives rise to the non-linear trend in the Raman frequency vs. strain plots for embedded graphenes regardless of size as reported previously [28]. Interestingly, in the case of simply supported graphene the transition from the stable phase to buckling instability is more abrupt and it is depicted in both experiments and MD simulations.

In **figure 3** we have plotted the dependence of the buckling wavelength on strain. The values were obtained through simple optical inspection of the corresponding simulation frames. At the critical point the wavelength was measured at 18.3 Å which is in good agreement with the value of 10–20 Å obtained from the Winkler analysis [28]. For larger intersurface distance $2Z_0$ the wavelengths increase. Under certain conditions (related to violation of inextensibility assumptions) it has been found that continuum theories cannot capture wrinkling phenomena at the nanoscale for suspended graphene over trenches[37] or under shear loadings[38]. However, herein, the wrinkling wavelength estimated by Winkler's model for the embedded graphene agrees well with the simulations. The main difference is that in the case of embedded graphene the inextensibility assumption is not violated due to the small magnitude of the out-of-plane deformations and the results suggest that the continuum model is valid since it is able to describe both the critical strain and the wrinkling characteristics.

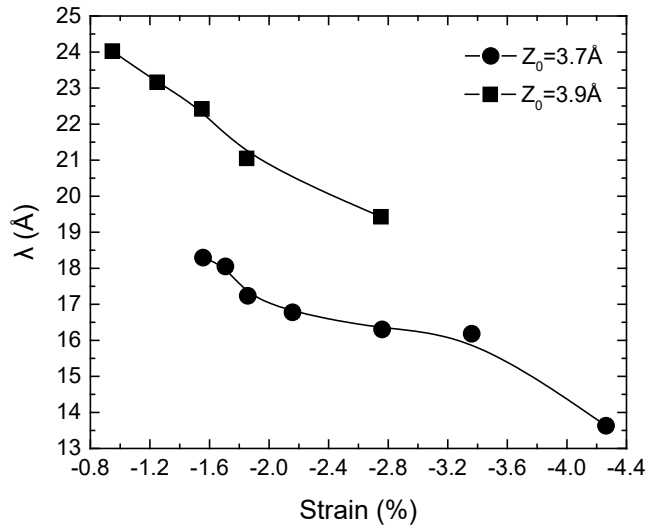

**Figure 3.** Wavelength of the sinusoidal forms dependence on strain for two intersurface distances $2Z_0$ (at $T = 300K$). At lower strain levels for each intersurface distances no periodic forms were observed. The lines are guides to the eye.

### *3.2 Simply supported graphene under compression*

Exfoliated graphene flakes were deposited on a PMMA/SU-8 polymer substrate using the scotch tape method. The samples were subjected to compressive deformations in a step-wise manner using a four-point-bending experimental setup. The Raman spectra for the 2D peak was recorded *in situ* at every strain level (see materials and methods section). An optical image of the examined flake and Raman spectra for the 2D peak for various strain levels are presented in **figures 4a** and **4b**, respectively. The length of the selected flake is about ~15 μm which should be sufficient to allow efficient stress transfer from the polymer to the flake[23].

In **figure 4c** the shift of the position of the 2D peak versus the applied strain is presented. As the compressive strain is applied phonon hardening is observed and the position of the 2D peak reaches a maximum value at an external strain of ~ −0.30%. This value corresponds to the critical strain to buckling, after which a significant phonon softening is observed since the graphene does not sustain further compression. Previous studies[22,27,28] have estimated by both analytical and experimental means that a 2D phonon shift rate of value ~50 cm$^{-1}$/% is required for 1:1 strain/stress transfer in graphene/PMMA systems. Thus, it is henceforth assumed that at the point of inflection, graphene is itself compressed to an identical strain of –0.30%. As mentioned earlier, it is worth noting that the critical buckling strain is half of the corresponding critical value for the formation of wrinkles in embedded graphene. Overall, the behavior up to failure is moderately non-linear whereas after failure the phonon softening is abrupt. As in the case of fully embedded graphene[28], the overall trend can be captured sufficiently by a second order polynomial curve as seen from **figure 4c** (ignoring the initial slack at 0% strain). In this regime the slope $\partial 2D/\partial \varepsilon$ is ~50.7 cm$^{-1}$/% which is close to the established shift of 55 cm$^{-1}$/%[27] for a laser line of 785 nm.

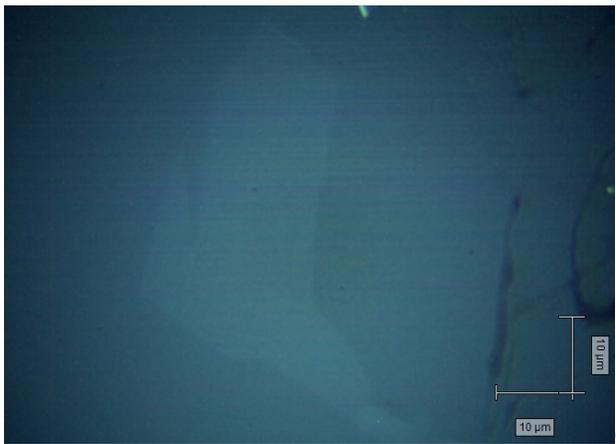
(a)

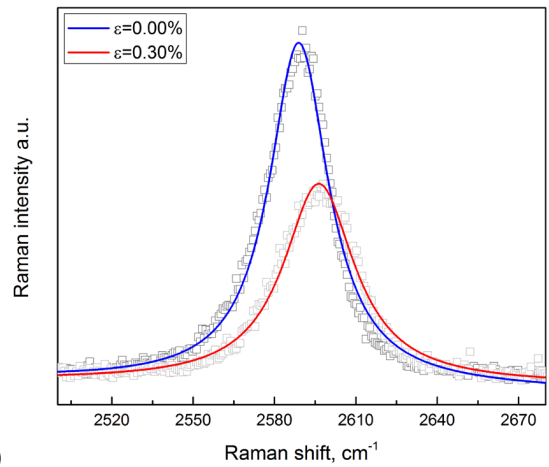
(b)

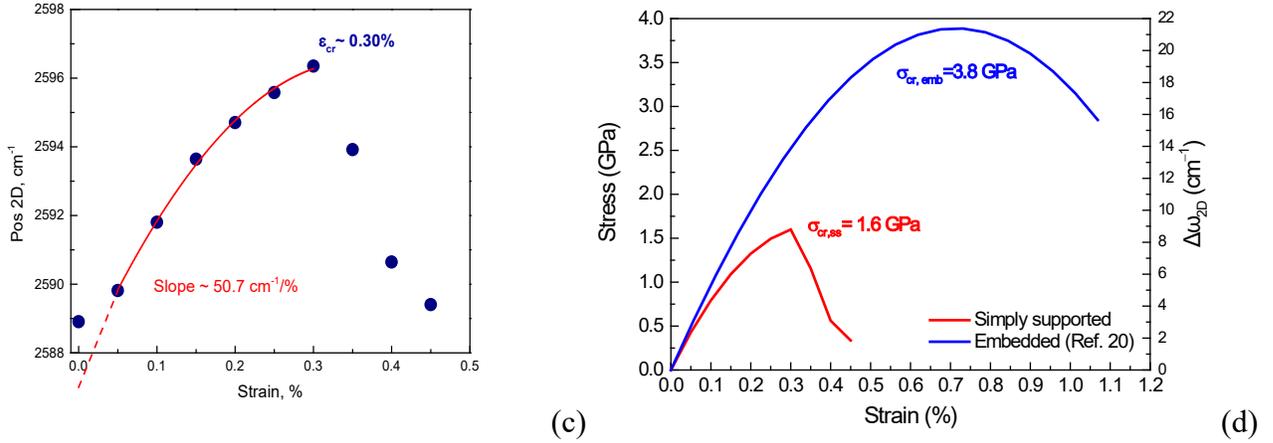

(c)   (d)

**Figure 4.** (a) Optical image of the examined single layer graphene flake. (b) Spectra of the 2D Raman peak for various strain levels. (c) The Pos(2D) versus the applied strain. (d) Stress–strain curves for the simply supported monolayer derived from spectroscopic data, and relative changes in 2D position frequencies with strain for simply supported and embedded graphene, with respect to the corresponding unstrained values.

Previously[27] we reported a method for converting spectroscopic data of the 2D Raman peak to values of axial stress and constructed stress–strain curves for a moderate strain range for embedded graphene under tension and compression. Using the estimated value of 5.5 cm$^{-1}$/GPa for a laser line of 785 nm, we derive the σ–ε curve for the simply supported graphene which is presented in **figure 4d**. Thus, we estimate that the critical compressive stress for the onset of buckling strength of a simply supported monolayer graphene is ~1.6 GPa and the initial Young's modulus 0.92 TPa. In **figures 4d** we have plotted in the same graph results for embedded and simply supported graphene to show the influence of the added upper layer of polymer which obviously enhances the resistance against buckling and the compressive strength of the single layer graphene. It is worth noting that the total shift of the position of the 2D peak of the simply supported graphene is about half of the embedded which is also depicted in the compressive strength. Thus, the added top layer doubles the resistance to buckling, in agreement with the DFT analysis where the value of $K_W$ was also doubled[28].

Other factors also affect the observed mechanical behavior. For example the polymer substrate in the simply supported case is not flat but contains a degree of roughness and therefore the graphene does not adhere perfectly well to the substrate. In the case of embedded graphene it is not known whether via the spin-coating procedure the resin fills the gaps of the bottom part of the substrate on which it is deposited. The morphology of the polymer surface should play a role as well, and this has been largely overlooked. Here we examine the effect of the interaction energy and of roughness on the critical strain for two forms of localized imperfections such as blocks and trenches.

For the interaction of the carbon atoms with the substrate we use a range of values between 1.0–20.0 meV/atom. To estimate critical strains for initiation of buckling we analyze several strain levels by producing average squared height plots that we present in **figure 5**. We repeat the process considering different interaction strengths with the substrate surface, specifically for 1.0, 6.7, 10.0 and 20.0 meV/atom. Based on these plots along with optical inspection of the images we obtain corresponding critical strains of –0.35, –0.90%, –1.18%, and 1.68%, respectively, and for a temperature of $T = 300$K. This result shows the strong dependence of the critical strain to buckling on the interaction strength, or equivalently on the material used as substrate for the simply supported graphene.

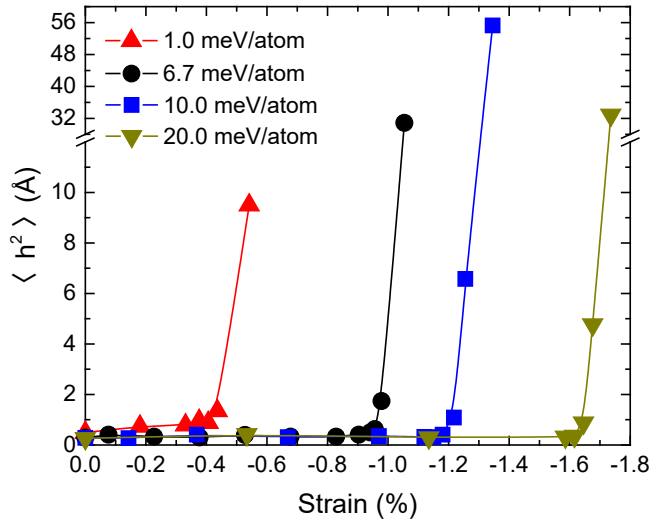

**Figure 5.** Squared height plots of the atomic distances from the substrate surface for various interaction strengths (at $T = 300$K). The lines are guides to the eye.

*3.3 Surface morphology*

To examine the effect of surface morphology on the compression behavior in the simply supported case, we have introduced localized artifacts in our previously ideal model in the form of a block positioned at the center of the interacting surface in the computational cell. We examine several cases that correspond to blocks of different height. In each case the block also interacts with the atoms of graphene with all its faces and with the same magnitude as the rest of the substrate surface, specifically by 6.7 meV/atom. The dimensions of the block are 50×100 Å$^2$ and with heights that vary.

In **figure 6** we show the case of a block with height 2 Å. The three frames are extracted from the simulation trajectory and correspond to compressive strains of 0.0%, –0.66%, and –0.75% from top to bottom. The positioning of the block can be easily identified in **figure 6a** as the lifted region colored in red and outlined by yellow. In the initially relaxed state, graphene adheres well on the block. As the compressive strain increases graphene begins to buckle at a certain threshold value. This value for buckling initiation has a strong dependence on the height of the block as shown clearly in **figure 7**. When the block height has reached ~6 Å, the graphene buckles under the slightest compression. We note that this dependence is not linear, but rather is reproduced by a parabola with a weak square dependence. In **figure 6b** it can be seen that buckling initiates on the side of the block due to reduced adhesion. This behavior is met consistently regardless of block height, except for very small heights of approximately less than 0.9 Å. For those cases it is almost equally probable for the graphene sample to buckle anywhere along the specimen regardless of block presence.

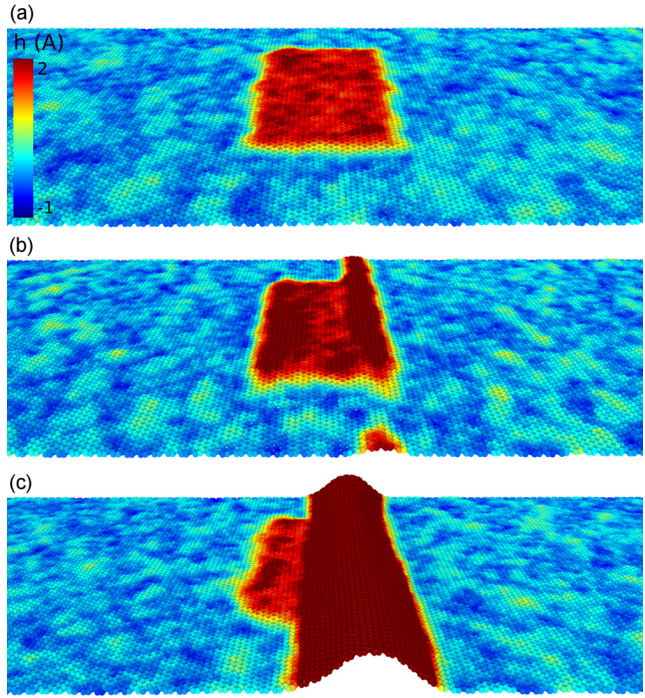

**Figure 6.** Localized buckling under compressive strain over a 50×100 Å$^2$ interacting block type imperfection with height 2 Å. (a) The block is underlaid in the red region outlined with yellow. The simulation frames correspond to strain levels of (a) 0.0%, (b) –0.66%, and (b) –0.75%. Atomic interactions with the substrate and block are 6.7 meV/atom (see text). The temperature is 300K. The color scale is chosen for clarity, not for accuracy.

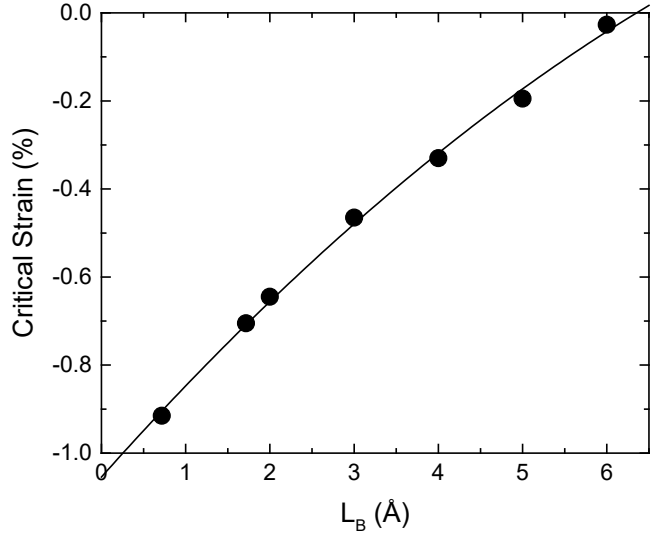

**Figure 7.** Critical strain level for initiation of buckling with respect to the height, $L_B$, of an interacting block (at $T$ = 300K). The line is a parabolic fit to the data (see text).

An alternate and interesting approach to model imperfections that make up surface roughness are extensive trenches. We introduce this type of imperfection in our previously ideal model by creating a region of controlled depth with respect to the rest of the substrate surface, positioned at the center of the interacting surface in the computational cell. The positioning of the trench can be easily identified in **figure 8a** as the strip region colored mainly in green, with a width of 100 Å and length from edge to edge (in the perpendicular direction) of the computational cell. We examine several cases that correspond to trenches of different depths. In each case the trench fully interacts with the atoms of graphene (bottom and side faces of the trench) and with the same magnitude (6.7 meV/atom).

In **figure 8** we show the case of a trench with a depth of 6 Å. Throughout the compression procedure several stages are encountered. This behavior in stages serves as a process that allows for graphene to better attach within the trench (extending closer to the side walls) prior to buckling and is observed for trenches deeper than 4 Å. Initially graphene is placed and relaxed over the surface and trench within the isothermal–isobaric ensemble as in the previous cases. When compressive strain is applied the graphene sheet remains suspended above the trench (**figure 8a**), then gradually starts entering into the trench and ultimately attaches to the bottom (**figure 8b**). In doing so the compression is partially relieved. We continue to apply compressive strain and as expected buckling is initiated, the onset of which is shown in **figure 8c**. Subtle difference can be identified between figures **8b and 8c**, specifically in case **8c** the red regions on the side of the trench are more pronounced, and the graphene spread tighter at the bottom of the trench (blue region). A further increment of external compressive strain of ~ –0.1% to –0.2% leads to the full buckling of graphene (an advanced stage of which is shown in **figure 8d**). For shallower trenches graphene readily attaches to the trench upon initial compression. Interestingly, for certain rather shallow trench depths after the formation of the buckle its inner front (with respect to the trench) reattaches to the substrate resulting in the propagation of the buckle away from the trench and with a non-reducible amplitude, similar to a solitonic wave. For the interacting surface of 6.7 meV/atom considered here, this occurs for depths of approximately 1–3 Å. Of course this may very well be a result of us using an ideal interacting surface, nevertheless it is conceivable that the phenomenon may persist even in realistic and smooth or periodic substrates, such as the ones fabricated by Kumaki *et al.*[39] based on a 1:2 (isotactic:syndiotactic) PMMA stereocomplex or the patterned substrates through relief structures fabricated by Bowden *et al.*[40]. However, a detailed study of this phenomenon is out of the scope of the present work.

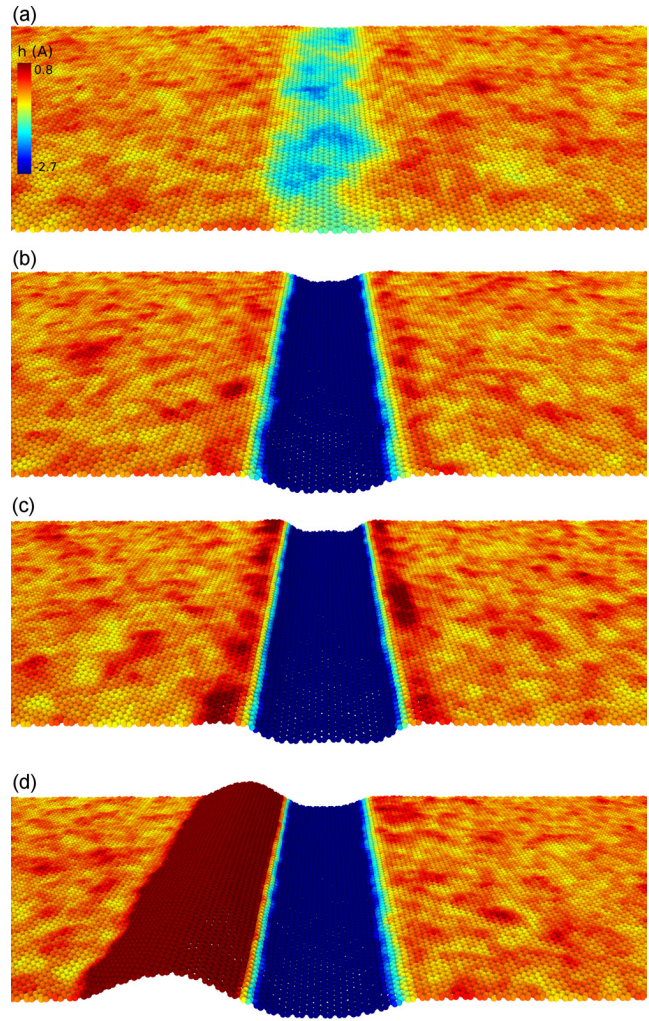

**Figure 8.** Stages during compression of graphene situated over a trench of depth 6 Å. (a) Initial stage in which the graphene sheet is suspended above the trench. (b) Upon compression the graphene sheet has gradually entered into the trench and ultimately attaches to the bottom of the trench. (c) Onset of buckling. Notice the intense red regions and the (now very narrow) yellow strip perimetric to the trench. (d) Advanced buckling state. The color scale is chosen for clarity, not for accuracy.

The strain levels can be defined by two separate ways depending on the stage taken as the zero strain level (graphene suspended over the trench, i.e. has not entered the trench which is actually the applied strain, or having entered the trench). In **figure 9** we show the critical strains for buckling with respect to the depth of the trench, $L_T$. Zero strain levels were taken both for graphene relaxed within the trench, and relaxed suspended over the trench. Only the former applies to the full range of examined trench heights. Critical strain levels corresponds to that just prior to the ultimate buckling.

When the critical values are taken with respect to graphene suspended over the trench we find a smooth dependence with trench depth up to a depth value of ~ 4 Å. After that, additional applied strain is required in order to overcome relaxation (partial relief of compression) induced when graphene attaches to the bottom of the trench. The other branch (zero strain level set with graphene attached to the bottom of the trench) exhibits larger absolute values since they are calculated with respect to a smaller initial length (denominator in strain). We note that the gap between the two branches depends on the sample's (computational cell) length expressed by the percentage of the graphene's length that enters into the trench. The decrease observed for larger depths results from reduction in effective length as more of graphene enters the trench and more tightly covers its side walls (further assisted by the applied compression). This does not occur in the case of block at least in the range of heights examined here.

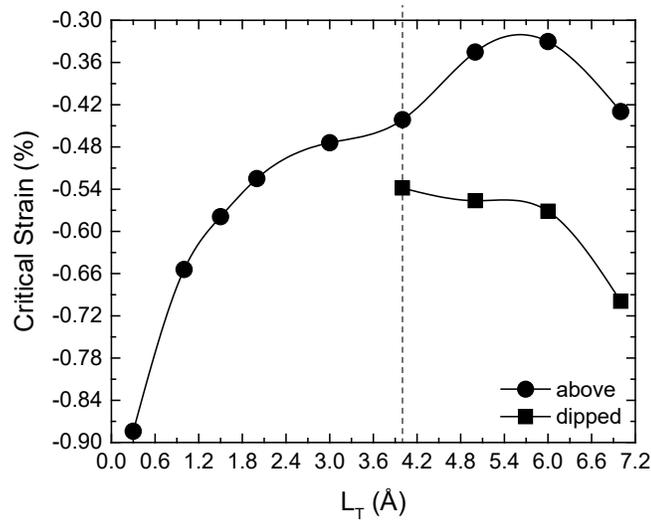

**Figure 9.** Critical strain level for initiation of buckling with respect to the depth, $L_T$, of an interacting trench (at $T$ = 300K). Zero strain levels are set either with graphene suspended over the trench, or attached (dipped) at the bottom of the trench. The lines are guides to the eye.

**4. Conclusions**

In the present study the mechanical behavior of single layer graphene simply-supported, as well as, fully embedded in PMMA was examined by means of experiment, theoretical analysis and molecular dynamics simulations. It was found that simply supported graphene wrinkles at a compressive strain of −0.30% which corresponds to a compressive strength of ~1.6 GPa. These values are about half of those found for the fully embedded case. A full comparison of the experimental results with the Winkler model, as well as, MD simulations is presented. The experimental critical strain is lower than the values obtained by the analytical modeling assuming ideal interacting surfaces. Possible deviations from ideal behavior was examined by introducing in the MD simulations the effect of polymer morphology (roughness, blocks and trenches) upon the critical strain to buckling. We found that the presence of localized surface imperfections decreases the critical strain to buckling in agreement with the experimental results.

**Acknowledgments**

This research has been co-financed by the European Research Council (ERC Advanced Grant 2013) via project no. 321124, "Tailor Graphene". Drs. Georgia Tsoukleri and John Parthenios are thanked for assistance and advice with the experiments. We acknowledge computation time provided by the Cy-Tera facility of the Cyprus Institute under the project CyTera. Finally, the authors acknowledge the financial support of the Graphene FET Flagship (''Graphene-Based Revolutions in ICT And Beyond''- Grant agreement no: 604391).

**References**

1   Lee, C., Wei, X. D., Kysar, J. W. & Hone, J. Measurement of the elastic properties and intrinsic strength of monolayer graphene. *Science* **321**, 385-388, doi:10.1126/science.1157996 (2008).
2   Galiotis, C., Frank, O., Koukaras, E. N. & Sfyris, D. Graphene Mechanics: Current Status and Perspectives. *Annual Review of Chemical and Biomolecular Engineering* **6**, 121-140, doi:10.1146/annurev-chembioeng-061114-123216 (2015).
3   Castro Neto, A. H., Guinea, F., Peres, N. M. R., Novoselov, K. S. & Geim, A. K. The electronic properties of graphene. *Rev Mod Phys* **81**, 109-162, doi:10.1103/RevModPhys.81.109 (2009).
4   Chen, J.-H., Jang, C., Xiao, S., Ishigami, M. & Fuhrer, M. S. Intrinsic and extrinsic performance limits of graphene devices on SiO2. *Nat Nano* **3**, 206-209, doi:http://www.nature.com/nnano/journal/v3/n4/suppinfo/nnano.2008.58_S1.html (2008).
5   Castro, E. V. *et al.* Limits on Charge Carrier Mobility in Suspended Graphene due to Flexural Phonons. *Phys. Rev. Lett.* **105**, 266601 (2010).
6   Mariani, E. & von Oppen, F. Temperature-dependent resistivity of suspended graphene. *Phys. Rev. B* **82**, 195403 (2010).
7   Shioya, H., Russo, S., Yamamoto, M., Craciun, M. F. & Tarucha, S. Electron States of Uniaxially Strained Graphene. *Nano Lett.*, doi:10.1021/acs.nanolett.5b03027 (2015).
8   Grima, J. N. *et al.* Tailoring Graphene to Achieve Negative Poisson's Ratio Properties. *Adv Mater* **27**, 1455-+, doi:10.1002/adma.201404106 (2015).
9   Koskinen, P. Graphene cardboard: From ripples to tunable metamaterial. *Appl Phys Lett* **104**, 101902, doi:doi:http://dx.doi.org/10.1063/1.4868125 (2014).
10  Zang, J. F. *et al.* Multifunctionality and control of the crumpling and unfolding of large-area graphene. *Nat Mater* **12**, 321-325, doi:10.1038/NMAT3542 (2013).

11  Polyzos, I. *et al.* Suspended monolayer graphene under true uniaxial deformation. *Nanoscale* **7**, 13033-13042, doi:10.1039/c5nr03072b (2015).
12  Cao, C., Feng, Y., Zang, J., López, G. P. & Zhao, X. Tunable lotus-leaf and rose-petal effects via graphene paper origami. *Extreme Mechanics Letters* **4**, 18-25, doi:http://dx.doi.org/10.1016/j.eml.2015.07.006 (2015).
13  Zhang, K. & Arroyo, M. Adhesion and friction control localized folding in supported graphene. *J Appl Phys* **113**, doi:Artn 193501
10.1063/1.4804265 (2013).
14  Al-Mulla, T., Qin, Z. & Buehler, M. J. Crumpling deformation regimes of monolayer graphene on substrate: A molecular mechanics study. *Journal of Physics Condensed Matter* **27**, doi:10.1088/0953-8984/27/34/345401 (2015).
15  Sfyris, D., Androulidakis, C. & Galiotis, C. Graphene resting on substrate: Closed form solutions for the perfect bonding and the delamination case. *International Journal of Solids and Structures* **71**, 219-232, doi:http://dx.doi.org/10.1016/j.ijsolstr.2015.06.024 (2015).
16  Sfyris, D., Koukaras, E. N., Pugno, N. & Galiotis, C. Graphene as a hexagonal 2-lattice: Evaluation of the in-plane material constants for the linear theory. A multiscale approach. *J. Appl. Phys.* **118**, doi:10.1063/1.4928464 (2015).
17  Jung, J. H., Bae, J., Moon, M.-W., Kim, K.-S. & Ihm, J. Numerical study on sequential period-doubling bifurcations of graphene wrinkles on a soft substrate. *Solid State Commun* **222**, 14-17, doi:http://dx.doi.org/10.1016/j.ssc.2015.08.020 (2015).
18  Aitken, Z. H. & Huang, R. Effects of mismatch strain and substrate surface corrugation on morphology of supported monolayer graphene. *J Appl Phys* **107**, doi:Artn 123531
10.1063/1.3437642 (2010).
19  Datta, D., Nadimpalli, S. P. V., Li, Y. & Shenoy, V. B. Effect of crack length and orientation on the mixed-mode fracture behavior of graphene. *Extreme Mechanics Letters* **5**, 10-17, doi:http://dx.doi.org/10.1016/j.eml.2015.08.005 (2015).
20  Gao, W., Liechti, K. M. & Huang, R. Wet adhesion of graphene. *Extreme Mechanics Letters* **3**, 130-140, doi:http://dx.doi.org/10.1016/j.eml.2015.04.003 (2015).
21  Jung, G., Qin, Z. & Buehler, M. J. Molecular mechanics of polycrystalline graphene with enhanced fracture toughness. *Extreme Mechanics Letters* **2**, 52-59, doi:http://dx.doi.org/10.1016/j.eml.2015.01.007 (2015).
22  Tsoukleri, G. *et al.* Subjecting a Graphene Monolayer to Tension and Compression. *Small* **5**, 2397-2402, doi:10.1002/smll.200900802 (2009).
23  Anagnostopoulos, G. *et al.* Stress Transfer Mechanisms at the Submicron Level for Graphene/Polymer Systems. *Acs Applied Materials & Interfaces* **7**, 4216-4223, doi:10.1021/am508482n (2015).
24  Jiang, T., Huang, R. & Zhu, Y. Interfacial Sliding and Buckling of Monolayer Graphene on a Stretchable Substrate. *Adv Funct Mater* **24**, 396-402, doi:10.1002/adfm.201301999 (2014).
25  Androulidakis, C. *et al.* Graphene flakes under controlled biaxial deformation. *Sci. Rep.* **5**, 18219, doi:10.1038/srep18219 (2015).
26  Vlattas, C. & Galiotis, C. Deformation-Behavior of Liquid-Crystal Polymer Fibers .1. Converting Spectroscopic Data into Mechanical Stress-Strain Curves in Tension and Compression. *Polymer* **35**, 2335-2347, doi:Doi 10.1016/0032-3861(94)90770-6 (1994).
27  Androulidakis, C. *et al.* Experimentally derived axial stress-strain relations for two-dimensional materials such as monolayer graphene. *Carbon* **81**, 322-328, doi:10.1016/j.carbon.2014.09.064 (2015).
28  Androulidakis, C. *et al.* Failure Processes in Embedded Monolayer Graphene under Axial Compression. *Sci. Rep.* **4**, doi:Artn 5271
10.1038/Srep05271 (2014).
29  Stuart, S. J., Tutein, A. B. & Harrison, J. A. A reactive potential for hydrocarbons with intermolecular interactions. *J. Chem. Phys.* **112**, 6472-6486, doi:doi:http://dx.doi.org/10.1063/1.481208 (2000).
30  Plimpton, S. Fast Parallel Algorithms for Short-Range Molecular-Dynamics. *J. Comput. Phys.* **117**, 1-19, doi:DOI 10.1006/jcph.1995.1039 (1995).

31  Jiang, J. W. The Strain Rate Effect on the Buckling of Single-Layer MoS2. *Sci. Rep.* **5**, doi:Artn 7814 10.1038/Srep07814 (2015).
32  Chen, M. Q. *et al.* Effects of grain size, temperature and strain rate on the mechanical properties of polycrystalline graphene - A molecular dynamics study. *Carbon* **85**, 135-146, doi:10.1016/j.carbon.2014.12.092 (2015).
33  Stukowski, A. Visualization and analysis of atomistic simulation data with OVITO–the Open Visualization Tool. *Modell. Simul. Mater. Sci. Eng.* **18**, 015012 (2010).
34  Lv, C. *et al.* Effect of Chemisorption on the Interfacial Bonding Characteristics of Graphene−Polymer Composites. *The Journal of Physical Chemistry C* **114**, 6588-6594, doi:10.1021/jp100110n (2010).
35  Sun, H. COMPASS:  An ab Initio Force-Field Optimized for Condensed-Phase ApplicationsOverview with Details on Alkane and Benzene Compounds. *The Journal of Physical Chemistry B* **102**, 7338-7364, doi:10.1021/jp980939v (1998).
36  Lin, F., Xiang, Y. & Shen, H. S. Buckling of Graphene Embedded in Polymer Matrix under Compression. *International Journal of Structural Stability and Dynamics* **15**, doi:10.1142/S0219455415400167 (2015).
37  Tapaszto, L. *et al.* Breakdown of continuum mechanics for nanometre-wavelength rippling of graphene. *Nat Phys* **8**, 739-742, doi:10.1038/Nphys2389 (2012).
38  Xiang, Y. & Shen, H.-S. Shear buckling of rippled graphene by molecular dynamics simulation. *Materials Today Communications* **3**, 149-155, doi:http://dx.doi.org/10.1016/j.mtcomm.2015.01.001 (2015).
39  Kumaki, J., Kawauchi, T., Okoshi, K., Kusanagi, H. & Yashima, E. Supramolecular Helical Structure of the Stereocomplex Composed of Complementary Isotactic and Syndiotactic Poly(methyl methacrylate)s as Revealed by Atomic Force Microscopy. *Angew. Chem. Int. Ed.* **46**, 5348-5351, doi:10.1002/anie.200700455 (2007).
40  Bowden, N., Brittain, S., Evans, A. G., Hutchinson, J. W. & Whitesides, G. M. Spontaneous formation of ordered structures in thin films of metals supported on an elastomeric polymer. *Nature* **393**, 146-149 (1998).